\documentclass[%
superscriptaddress,
longbibliography,
amsmath,amssymb,
prl,
twocolumn,
]{revtex4-1}
\usepackage[export]{adjustbox}
\usepackage{mathtools}
\usepackage{graphicx}% Include figure filessn
\usepackage{color}% Include figure files
\usepackage{xcolor}% Include figure files
\usepackage{wasysym}% Include figure files
\usepackage{dcolumn}% Align table columns on decimal point
\usepackage{bm}% bold math
\usepackage{hyperref}% add hypertext capabilities
\usepackage[mathscr]{euscript}
\graphicspath{{fig/}}
\definecolor{mygreen}{rgb}{0,0.5,0}
\definecolor{mybrown}{rgb}{0.65,0.16,0.16}
\newcommand{\colg}[1]{\textcolor{mygreen}{#1}}

\newcommand{\colr}[1]{\textcolor{red}{#1}}
\newcommand{\colm}[1]{\textcolor{magenta}{#1}}

\newcommand{\colc}[1]{\textcolor{cyan}{#1}}
\newcommand{\colbr}[1]{\textcolor{mybrown}{#1}}

\def\beq {\begin{equation}}
\def\eeq {\end{equation}}
\def\beqa {\begin{eqnarray}}
\def\eeqa {\end{eqnarray}}
\def \bnum {\begin{enumerate}}
\def \enum {\end{enumerate}}
\def\bi {\begin{itemize}}
\def\ei {\end{itemize}}
\def \bdes {\begin{description}}
\def \edes {\end{description}}

\def\rel {R_{\lambda}}
\def\pel {{Pe}_{\lambda}}

\def\etak{\eta_{K}}
\def\dho{\partial}
\def\la {\langle}
\def\ra {\rangle}

\def\mbf {\mathbf}
\def\dx {\Delta}
\def\zith{\zeta^\theta_p}
\def\zitwo{\zeta^\theta_2}
\def\zifour{\zeta^\theta_4}
\def\zisix{\zeta^\theta_6}

\def\psif{\psi^\theta_4}
\def\psih{\psi^\theta_6}

\def\zinf{\zeta^\theta_\infty}

\def\drtheta{\delta_r \theta}
\def\ux{\mbf x}
\def\ur{\mbf r}

\def\nabv{\mbf\nabla}
\def\normr{r/\eta_K}
\def\pdf{\mathcal{P}}
\def\thetarms{\theta_{\textrm{rms}}}

\def\scgx{\dho\theta/\dho x}
\def\scglarge{\theta_{rms}/\eta_K}
\begin{document}
%\setlength{\abovedisplayskip}{3pt}
%\setlength{\belowdisplayskip}{3pt}
%\preprint{APS/123-QED}
\title{Steep cliffs and saturated exponents in three dimensional scalar turbulence}
%%%%%%%%%%%%%%%%%%%%%%%%%%
\author{Kartik P. Iyer}
\email{kartik.iyer@nyu.edu}
\affiliation{Tandon School of Engineering, New York University, New York, NY 11201, USA}
%%%%%%%%%%%%%%%%%%%%%%%%%%
\author{J\"{o}rg Schumacher}
\affiliation{Tandon School of Engineering, New York University, New York, NY 11201, USA}
\address{Institut f\"ur Thermo-und Fluiddynamik, Technische Universit\"at Ilmenau, Postfach 100565, D-98684 Ilmenau, Germany}
%%%%%%%%%%%%%%%%%%%%%%%%%%
\author{Katepalli R Sreenivasan}
\affiliation{Tandon School of Engineering, New York University, New York, NY 11201, USA}
\address{Department of Physics and the Courant Institute of Mathematical Sciences, New York, NY 10012, USA}
%%%%%%%%%%%%%%%%%%%%%%%%%%
\author{P K Yeung}
\address{School of Aerospace and Mechanical Engineering, Georgia Institute of Technology, Atlanta, GA 30332, USA}

\date{\today}% It is always \today, today,
\begin{abstract}
The intermittency of a passive scalar advected by three-dimensional Navier-Stokes turbulence at a Taylor-scale Reynolds number of $650$ is studied using direct numerical simulations on a $4096^3$ grid; the Schmidt number is unity. By measuring scalar increment moments of high orders, while ensuring statistical convergence, we provide unambiguous evidence that the scaling exponents saturate to $1.2$ for moment order beyond about $12$, indicating that scalar intermittency is dominated by the most singular shock-like cliffs in the scalar field. We show that the fractal dimension of the spatial support of steep cliffs is about $1.8$, whose sum with the saturation exponent value of $1.2$ adds up to the space dimension of $3$, thus demonstrating a deep connection between the geometry and statistics in turbulent scalar mixing. The anomaly for the fourth and sixth order moments is comparable to that in the Kraichnan model for the roughness exponent of $2/3$.
\end{abstract}
\pacs{Valid PACS appear here}% PACS, the Physics and Astronomy

\maketitle

%%%%START%%%%%%%%%%%%%
Non-Gaussianity and intermittency are the norm in non-equilibrium statistical physics \cite{BHP98}, astrophysics \cite{bruno2013,zel89}, physical oceanography \cite{Mashayek17}, outdoor fires \cite{arfm01}, and many other applications. The basic characteristics of intermittent systems, namely the intense and sporadic fluctuations of the small scale, are not captured by classical mean field theories. Two important examples of intermittent fluid systems are three-dimensional (3D) Navier-Stokes (NS) turbulence \cite{Fri95,SA97} and 3D scalar turbulence \cite{KRS91,ZW00}---a 
short phrase for passive scalars mixed by NS turbulence. Scalar turbulence describes how a passive concentration field is transported by an advecting flow, generating scalar fluctuations on progressively wider ranges of scales and enhancing the rate of mixing of the scalar with its surroundings \cite{MY.II,SS2000}. The local concentration gradients are accompanied by large bursts of fluctuations at scales smaller than the stirring scales, and are increasingly amplified down to a scale that is quenched by molecular diffusivity. Much progress in understanding intermittency has been made in related problems of forced Burgers turbulence \cite{BMP95,mitra05,bec07} and the Kraichnan model \cite{RK68} for a scalar advected by a synthetic velocity field with no memory with respect to time; see, e.g., Refs.\ \cite{RK94,VY97,cher97,balkovsky98,CLMV2000,GF01,kalda08}. The existence of finite-time correlations of the advecting flow in 3D NS turbulence is one reason why theoretical progress has been slow; the high spatial and temporal resolution requirements put considerable strain on empirical work, which has slowed progress on that front. A large number of experimental and numerical efforts continue to be made on understanding scalar intermittency, e.g., Refs.~\cite{Antonia84,MSF90,EL99,Moisy2001,skrbek02,WG04,GW04,WG06,Gotoh15}, but the influence of large-scale quasi-discontinuities, which are denoted as {\em cliffs} or {\em fronts}, has remained an open question.
%%%%%%%%%%%%%%%%%%
\begin{figure}
\begin{minipage}[t]{0.5\textwidth}
\includegraphics [width=0.9\textwidth]{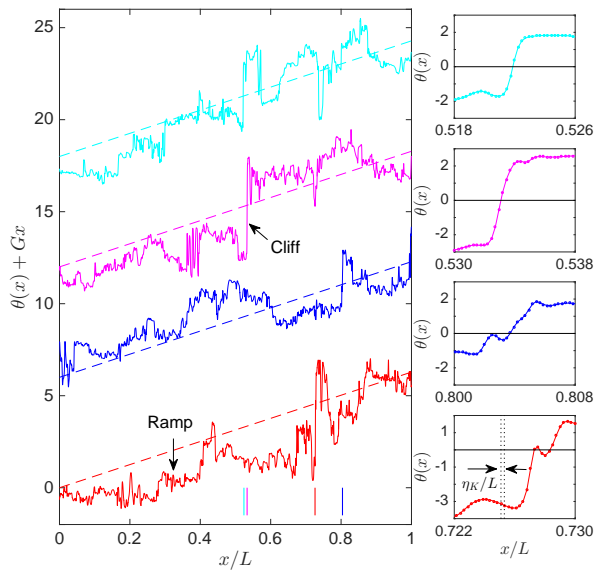}
\end{minipage}
\vspace{-0.9cm}
\protect\caption{Ramp-cliff structures in a scalar field, $\Theta \equiv \theta + Gx$, here $\theta$ is the scalar fluctuation and $(G,0,0)$ is the mean gradient, at $\rel=650$ and the Schmidt number $Sc\equiv \nu/D=1$, where $\nu$ is the kinematic viscosity of the fluid and $D$ is the scalar diffusivity. $L$ is the size of the computational cube in one direction. The main figure to the left plots four 1D profiles of $\Theta$ in the $x$-direction, along which the mean gradient is imposed. Examples for ramps and cliffs are indicated by arrows as well as the mean scalar concentration profile (dashed lines). Profiles are shifted in steps of $5$ units with respect to each other for clarity.  The vertical solid lines indicate the spatial positions for the magnifications of the scalar fluctuation profiles plotted to the right. Grid resolution and Kolmogorov length $\eta_K$ are indicated.}
\label{fig1.fig}
\end{figure}
%%%%%%%%%%%%%%%%%%%

In this Letter, we report the precise quantification of the small-scale intermittency of a statistically stationary scalar turbulence, advected by 3D high-Reynolds-number, isotropic Navier-Stokes turbulence in a high-resolution direct numerical simulations (DNS). We connect the statistical footprints of the well-mixed scalar regions known as {\em ramps}, e.g., Refs.\ \cite{Mestayer76,gibson77,sab_79,antonia79,HS94,pumir94,MW98}, to the steep cliff regions (see Fig.~\ref{fig1.fig} and later in the text). The scaling exponents $\zith$ of the scalar correlations, which will be defined further below, saturate for moment orders above about $12$, to a constant $\zeta^{\theta}_{\infty}$ confirming that the almost-shock-like steep scalar fronts characterize scalar intermittency in 3D NS turbulence. We will also show that the spatial support of the cliffs with a fractal dimension of $D_F=1.8$ combines with the saturated scaling exponent $\zeta^{\theta}_{\infty}$ to the space dimension, yielding the result $\zeta_{\infty}^{\theta}+D_F=3$, where 3 is the space dimension, thus demonstrating the intimate link between the geometry and statistics in turbulent passive scalar mixing.

\noindent{\em Numerical simulations.} We use data from pseudo-spectral DNS of isotropic turbulence, computed using $4096^3$ mesh points in a periodic box of size $L$. A statistically steady state was obtained by forcing the low Fourier modes of the velocity field. The Taylor-scale Reynolds number $\rel = 650$; since the Schmidt number $Sc = 1$, the Taylor-scale P\'{e}clet number $\pel \equiv \rel Sc = 650$. The grid resolution $\dx/\etak = 1.1$, $\dx$ being the grid spacing and $\etak$ the Kolmogorov length scale. This resolution may not be adequate for capturing all aspects of the dissipative features, but is deemed quite adequate for assessing inertial range properties. The passive scalar $(\Theta)$ is evolved using the diffusion-advection equation in the presence of a uniform mean gradient $\mbf{G} \equiv (G,0,0)$ along the $x$-direction, where $G \ne 0$ is a constant, such that $\Theta = \theta + Gx$, $\theta$ here is the scalar fluctuation \cite{KI14}.  Averages over ten large-eddy turnover times were used in the analysis. In total, we use 210 trillion data samples to compute the statistics. For further details on the simulation and velocity field statistics, see Refs.~\cite{YDS2012,KI15}. 

\noindent{\em Ramps and cliffs.} The scalar signal organizes itself into conspicuous patterns as shown in Fig.~\ref{fig1.fig}, consisting of two distinctive features: (i) ramp regions where the total scalar gradient $\nabv\theta+\mbf{G}$ is of the order of $G$; and (ii) high gradient cliffs which are interspersed between ramps. The small figures on the right demonstrate clearly that the scalar increment, $\drtheta \equiv \theta(\ux+\ur)-\theta(\ux)$, can jump by the order $GL$ over $r \equiv |\ur|$ that is just a few multiples of the Kolmogorov scale $\eta_K$ (which is also the smallest dynamically significant scale in the scalar field). The ramp-cliff structure are connected to the mean scalar gradient in the present DNS, and are known to cause the breakdown of local isotropy in the scalar field \cite{KRS91}. The cliffs are caused by the action of large scales in the scalar field, even in the absence of a mean gradient \cite{KRS91,CK98}. The generic existence of scalar cliffs in turbulence suggests that these local spatial barriers to scalar mixing have a significant impact on scalar intermittency.
%%%%%%%%%%%%
\begin{figure}
\begin{minipage}[t]{0.5\textwidth}
\includegraphics [width=0.9\textwidth]{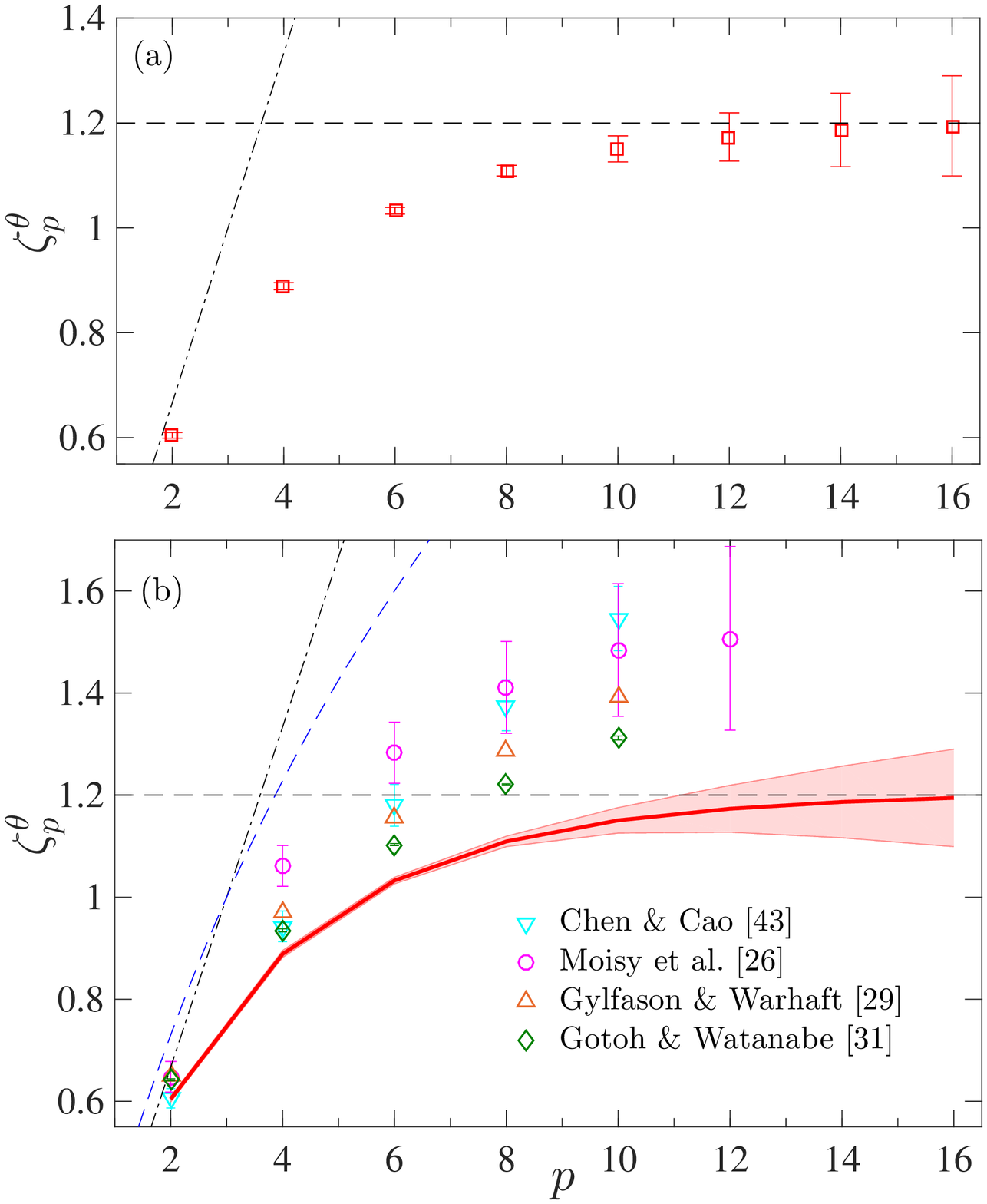}
\end{minipage}
\vspace{-0.7cm}
\protect\caption{
Scalar increment exponent $\zith$ {\it{vs}} moment order $p$. (a) present DNS: $\pel = 650$: dashed line at saturation exponent, $\zinf = 1.2$. Error bars indicate $95\%$ confidence interval. (b) Comparison of present DNS (shaded region) with previous results: $(\colc{\triangledown})$ $\pel = 220$ \cite{CC97}; $(\colm{\ocircle})$ $\pel = 280$ \cite{Moisy2001}; $(\colbr{\triangle})$ $\pel = 396$ \cite{GW04}; $(\colg{\lozenge})$ $\pel = 580$ \cite{Gotoh15}. Dash-dotted line shows normal scaling $\zith=p/3$; dashed line is the model of Ref.~\cite{EL99}. 
%The intersection of the two asymptotes intersect at $p=3.6$, whose significance is discussed briefly in the text.
}
\label{fig2.fig}
\end{figure}
%%%%%%%%%%%%

\noindent{\em Passive scalar increments.} In order to assess scalar intermittency, we define the $p$th order scalar structure function, $S^p_\theta(\ur) \equiv \la (\drtheta)^p \ra$, where $\la \cdot \ra$ denotes space/time averages. Due to the anisotropic mean scalar gradient, $S^p_\theta(\ur)$ depends on the separation vector $\ur$. However, the isotropic sector $\la(\drtheta)^p\ra_{0}$ 
extracted from the SO(3) decomposition \cite{Kurien2001,bp05} of $S^p_\theta(\ur)$, is only a function of scalar separation $r$ \cite{KI17b}. For scale $r$ in the inertial range, $\eta_K \ll r \ll \ell$, where $\ell$ is the integral scale of the velocity field, taken as $\normr \in [30,300]$ \cite{KI15}, $\la(\drtheta)^p\ra_{0}$ are found to follow power laws, $\la(\drtheta)^p\ra_{0} \sim r^{\zith}$, where $\zith$ denote the $p$th order scaling exponents. The higher order exponents are determined using extended self-similarity (ESS) \cite{ESS93}, by plotting $\la(\drtheta)^p\ra_{0}$ against $\la(\drtheta)^2\ra_{0}$ for $p > 2$. We have verified that estimating $\zith$ using the method of local slopes, e.g. \cite{Gotoh15}, or compensated structure functions, e.g. \cite{Moisy2001}, yield results consistent with the ESS results. 

%%%%%%%%%%%%%%%%%%%%%%%%%%
\begin{figure}
\begin{minipage}[t]{0.5\textwidth}
\includegraphics [width=0.8\textwidth]{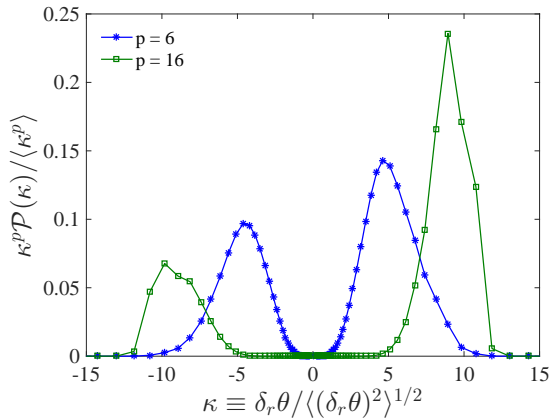}
\end{minipage}
\vspace{-0.7cm}
\protect\caption{Integrands of scalar increment moments ($\pdf(\cdot)$ denotes PDF of $(\cdot)$) as a function of the scalar increments, for orders $6$ and $16$ at $\normr = 55$ (lower end of the inertial range); moments of orders up to $20$ converge as well (and confirm the saturation of exponents) but are not shown here. The integrands are normalized by respective moments such that the area under each curve is unity.
}
\label{fig3.fig}
\end{figure}
%%%%%%%%%%%%%%%%%

\noindent The scaling exponents $\zith$ are plotted against moment order $p$ at $\pel=650$ in Fig.~\ref{fig2.fig}(a). The exponents saturate to $\zinf = 1.2$, indicated by the horizontal line, for $p \geq 12$. This is the clearest indication that the scalar fluctuations are limited in magnitude only by the largest allowable gradients in the field (largest temperature difference divided by the smallest length scale). The $\zinf=1.2$ curve intersects the normal scaling curve at $p=3.6$. In some sense, it is possible that this represents the situation for infinitely large $\pel$.
%Saturation of exponents implies that the anomaly 
%from normal scaling
%(indicated in Fig.~\ref{fig2.fig} by the dash-dot line),
%$\psith \equiv (p/2)\zitwo-\zith$, increases with $p$,
%showing that the scalar field is increasingly intermittent (see also Fig.~\ref{fig6.fig} later in the text).
Figure \ref{fig2.fig}(b) compares the present exponents with previous results for lower $\pel$. While our data robustly confirm that the exponents saturate, it is hard to reach a similar unambiguous conclusion from the previous results in the literature \cite{RCBC96,CC97,Moisy2001,skrbek02,WG04,GW04,Gotoh15}.
%%%%%%%%%%%%%%%%%
\begin{figure}
\begin{minipage}[t]{0.5\textwidth}
\includegraphics [width=0.8\textwidth]{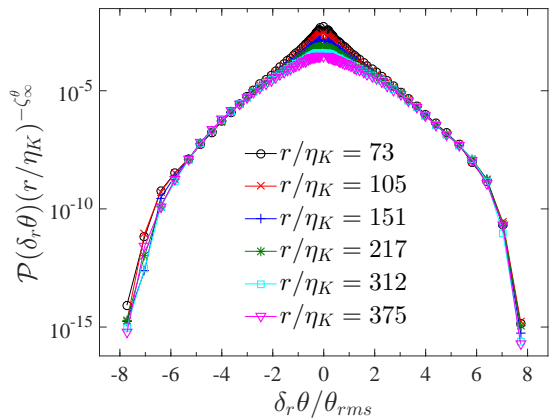}
\end{minipage}
\vspace{-0.7cm}
\protect\caption{PDF of scalar increments across the inertial range, multiplied by $r^{-\zinf}$, where $\zinf$ is the saturation exponent  (Fig.~\ref{fig2.fig}). The PDF tails collapse, confirming saturation of exponents.}
\label{fig4.fig}
\end{figure}
%%%%%%%%%%%%%%%%%

The statistical convergence of the moments of order $p$ up to $20$ was confirmed by the 
rapid decay of the moment integrands, $(\drtheta)^p \pdf(\drtheta)$, 
where $\pdf$ denotes the probability density function (PDF). 
%The integrands of moment orders $6$, $16$, $18$ and $20$
The integrands of moment orders $6$ and $16$
are shown in Fig.~\ref{fig3.fig}, each for $r$ in the low end of the inertial range. 
The integrands peak before the tail contributions decay, ensuring statistical convergence 
of the moments. Saturation of exponents at higher orders implies that, 
for scalar jumps $|\drtheta| \gtrsim \thetarms$, $\pdf(\drtheta) \propto r^{\zinf}$, 
where $\thetarms \equiv \sqrt{\la{\theta}^2\ra}$ \cite{CLMV2000,CLMV01}. 
Figure \ref{fig4.fig} verifies that this is indeed the case, 
with $\pdf(\drtheta) r^{-\zinf}$ collapsing for $|\drtheta| \ge 3\thetarms$, 
for all inertial separations. The inference is that the 
saturation of exponents arises because of the dominance of the 
high order moments by features that do not change with scale, 
suggesting that the gradients are of the order $\theta_{rms}/\eta_K$.

%%%%%%%%%%%%%%
\begin{figure}
\begin{minipage}[t]{0.5\textwidth}
\includegraphics [width=0.8\textwidth]{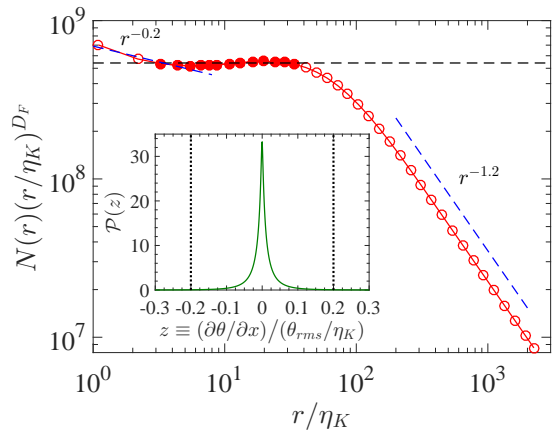}
\end{minipage}
\vspace{-0.7cm}
\protect\caption{
Log-log plot of the number $N(r)$ of cubes of side $r$ containing the steepest fronts {\it{vs}} size $r$. The ordinate is compensated by $r^{D_F}$, where $D_F = 1.8$ is the fractal co-dimension of the fronts. The plateau region $(\colr{\CIRCLE})$ corresponds to the scaling $r^{-D_F}$, indicated by the horizontal dashed line. At the smallest and largest $r$, the dimension of the fronts is $2$ (corresponding to flat fronts) and $3$ (which is the Euclidean dimension of the flow), respectively. Inset shows the PDF of the normalized gradient $z$,
$|z| > 0.2$ (dotted lines) is used to calculate $N(r)$.   
}
\label{fig5.fig}
\end{figure}
%%%%%%%% END %%%%%%%%%%%%%%

\noindent{\em Spatial support of cliffs.} We now turn to quantifying the dimension of the spatial support of the cliffs where strong scalar gradients tend to concentrate in sharp fronts (Fig.~\ref{fig1.fig}). The dimension of such fronts is estimated by the spatial support of regions of the strongest gradients of ${\cal O}(\scglarge)$ with cubes of edge size $r$, and counting their respective number $N(r)$ for different $r$. As shown in the inset of Fig.~\ref{fig5.fig}, gradients greater than $20\%$ of $\scglarge$ (marked by dotted lines) corresponding to $5 \sqrt{\la (\scgx)^2 \ra}$, are used to determine $N(r)$. We chose the threshold of $20\%$ as a good representative of gradients of the order $\theta_{rms}/\eta_K$ occurring with low probability (see inset to Fig.\ 5). [The use of a somewhat different threshold alters the scaling range in Fig.\ 5 but does not alter the dimension itself.] The plot of $N(r)$ {\it{vs}} $r$ for such fronts shown in the main body of Fig.~\ref{fig5.fig}, is compensated by $r^{1.8}$ (see below for the rationale), and has three scaling regimes: (i) at the smallest scales, a slope of $-2$ which corresponds to flat fronts; (ii) at $\normr \in [4,30]$, for which the slope from the least-squares fit is $D_F = 1.79 \pm 0.01$, corresponding to the spatial subset that supports the steep fronts in the scalar field; (iii) at the largest scales, the slope is $-3$ which corresponds to the Euclidean dimension of the flow. We confirm, for the first time in 3D NS flows, that the saturation exponent $\zinf$ and the box counting dimension of the steep fronts $D_F$ are related to the space dimension, $d = 3$, as 
\beq
\label{zinf.eq}
\zinf + D_F = d \;. 
\eeq
The confirmation of this relation in Navier-Stokes turbulence is remarkable since it directly connects a property of the highly intermittent statistics of the scalar to the spatial geometry of mixing barriers in the flow.

%%%%%%%%%%%%%%
\begin{figure}
\begin{minipage}[t]{0.5\textwidth}
\includegraphics [width=0.8\textwidth]{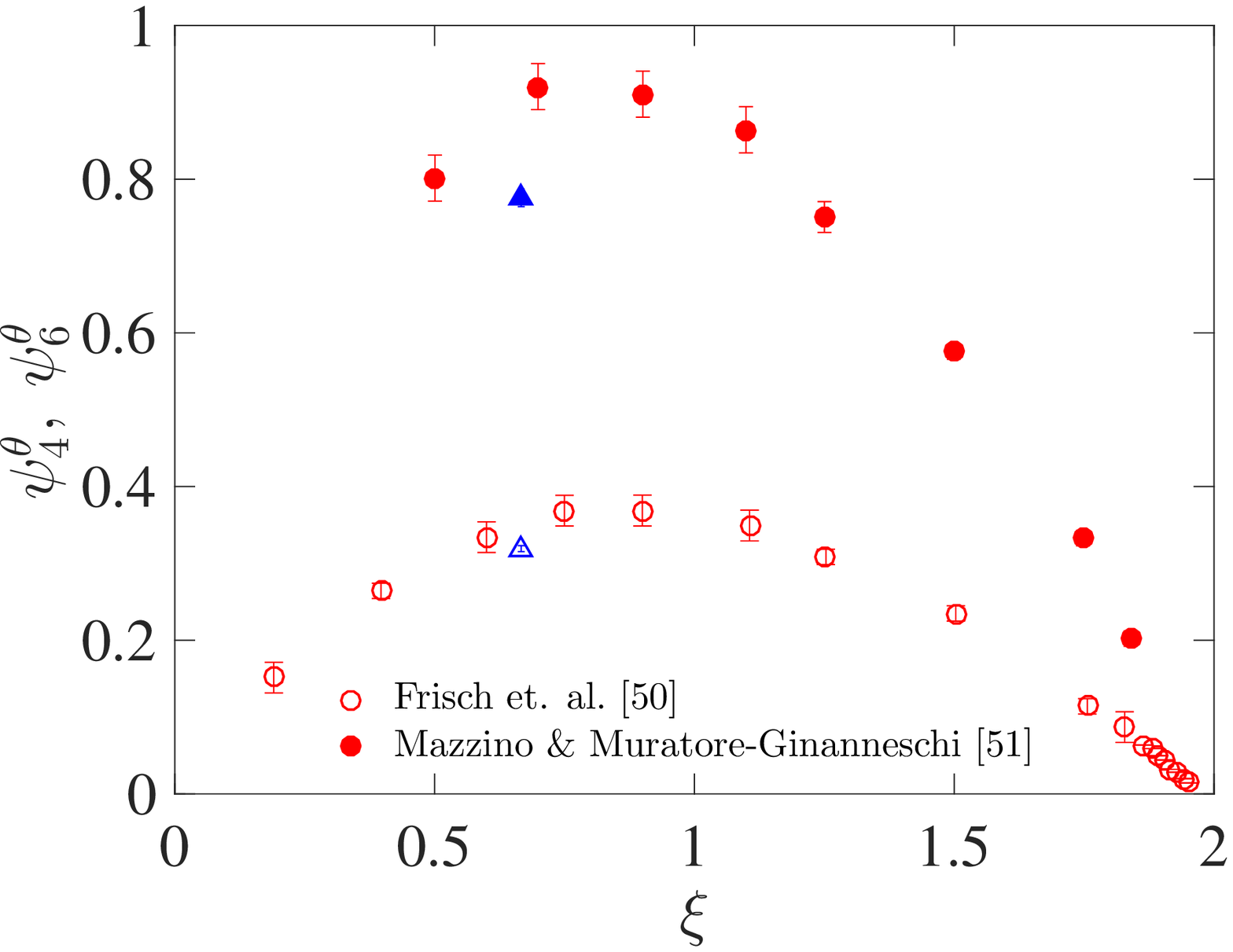}
\end{minipage}
\vspace{-0.7cm}
\protect\caption{Flatness anomaly $(\psif = 2\zitwo - \zifour)$ (open symbols) and hyper-flatness anomaly $(\psih = 3\zitwo - \zisix)$ (closed symbols) for the scalar {\it{vs}} flow roughness $\xi$.
Circles correspond to 3D Kraichnan model, $(\colr{\Circle})$ \cite{UF98} and $(\colr{\CIRCLE})$ \cite{MMG00}, while triangles are for 3D NS flow at $\pel=650$, with the roughness parameter $\xi = 2/3$.
}
\label{fig6.fig}
\end{figure}

\noindent{\em Correspondence with the Kraichnan flow.} The anomaly in the passive scalar field advected by a 3D NS flow at high-$\rel$ is comparable for orders $4$ and $6$ to that advected by the $\delta$-correlated 3D Kraichnan model \cite{UF98,MMG00}, as seen in Fig.~\ref{fig6.fig} where we quantify the degree of anomaly of exponents against the roughness parameter of the flow (which varies between $0$ and $2$ for Kraichnan model and is $2/3$ for the Navier-Stokes turbulence). This observed agreement for moderate-order moments is plausible because, in a high-$\rel$ NS flow, the small scales evolve with temporal rapidity, thereby approach the Kraichnan limit of a flow without memory. Scalar exponents for the Kraichnan model saturate at different values for different roughness parameters \cite{CLMV2000,CLMV01}, and the observed correspondence with the 3D NS results may not hold for high-order moments. 
%It appears from the intersection of the asymptotes in 
%Fig.\ 2a at 2.7, which may well be the crossover moment order marking the saturation.
%In the 3D Kraichnan model, it is shown in Ref.~\cite{balkovsky98} that the saturation crossover occurs at $p=3$.
%The present data suggest that a, a similar behavior is expected to occur where the  
%$\zinf=1.2$ scaling intersects the normal scaling curve at order $p = 3.6$ in
%Fig.~\ref{fig2.fig}(a), which may
%well be the crossover order marking the saturation, at asymptotic $\pel$ 
%(see also Ref.~\cite{balkovsky98}).

\noindent{\em Conclusions.} Our conclusive result here is that in a scalar field advected by 3D NS turbulence, the exponents $\zith$ saturate to $\zinf$ at large orders and that the saturation exponent is connected to the fractal dimension of the steep fronts. We do not expect that $\zinf$ itself to be universal \cite{chertkov96,SS96,LP09}, but the fact that scalar exponents saturate in 3D NS flows can have important consequences. For instance, the minimum H\"{o}lder exponent of $\theta$, $h_{min}^{\theta} \coloneqq \lim_{p \to \infty} \zith/p =  \lim_{p \to \infty} \zinf/p = 0$, which implies that shock-like quasi-discontinuities, or steep fronts, characterize the large gradients of the scalar field, somewhat reminiscent of 1D Burger's flow. However, while the Burger's flow displays a bi-scaling behavior, the lower order scalar exponents appear to have a quadratic dependence on the order, similar to that derived for scalar advection in high-dimensional Kraichnan model \cite{balkovsky98}. These results should have important implications for further theoretical understanding and modeling of scalar turbulence.
%%%%%%%% END %%%%%%%%%%%%%%
\section{Acknowledgments}
The computations and data analyses reported in this paper were performed using advanced computational facilities provided by the Texas Advanced Computation Center (TACC) under the XSEDE program supported by NSF.  The datasets used were originally generated using supercomputing resources at the Oak Ridge Leadership Computing Facility at the US Department of Energy Oak Ridge National Laboratory. The work of JS was supported by the Tandon School of Engineering at New York University and Grant No. SCHU 1410/19-1 of the Deutsche Forschungsgemeinschaft.
%%%%%%%%%%%%%%%% THE END %%%%%%%%%%%
\bibliography{zebib}
%%%%%%%%%%%%%%%%%%%%%%%%%%%%%
\end{document}